\documentclass[aps,showpacs,preprintnumbers,twocolumn,amsmath,amssymb,superscriptaddress,floatfix,nofootinbib]{revtex4-1}
\usepackage{lipsum}
\usepackage{graphicx}
\graphicspath{{final/}}
\usepackage{epsfig}
\usepackage{epstopdf}
\usepackage{hyperref}
\usepackage{amsmath}
\usepackage{amsfonts}
\usepackage{amssymb}

\newcommand{\im}{{\rm Im}}

\newcommand{\mlamc}{M_{\Lambda_c^+}}
\newcommand{\mev}{{\rm MeV}}

\DeclareMathSizes{12}{12}{12}{12}
\usepackage[section]{placeins}
\usepackage{slashed}

\begin{document}
\title{Production of $N^*(1535)$ and $N^*(1650)$ in
$\Lambda_c\rightarrow\bar{K}^0\eta p$ $(\pi N)$ decay}
\date{\today}

\author{R. Pavao}
\email{rpavao@ific.uv.es}

\author{S. Sakai}

\author{E.~Oset}
\affiliation{Departamento de
F\'{\i}sica Te\'orica and IFIC, Centro Mixto Universidad de
Valencia-CSIC Institutos de Investigaci\'on de Paterna, Aptdo.22085,
46071 Valencia, Spain}

\begin{abstract} 
 In order to study the properties of the $N^*$(1535) and $N^*$(1650) we
 calculate the mass distributions of $M B$ in the $\Lambda_c \rightarrow
 \bar{K}^0 M B$ decay, with $MB=\pi N(I=1/2),\eta p$ and
 $K\Sigma(I=1/2)$.
 We do this by calculating the tree-level and
 loop
 contributions, mixing pseudoscalar-baryon and vector-baryon channels
 using the local hidden gauge formalism.
 The
 loop
 contributions for each channel are calculated using the chiral unitary
 approach. We observe that for the $\eta N$ mass distribution only the
 $N^*$(1535) is seen, with the $N^*$(1650) contributing to the width of
 the curve, but for the  $\pi N$ mass distribution both resonances are
 clearly visible.
 In the case of $MB=K\Sigma$, we found that the strength of the $K\Sigma$ mass
 distribution is smaller than
 that of the mass distributions of the $\pi N$ and $\eta p$ in
 the $\Lambda_c^+\rightarrow\bar{K}^0\pi N$ and
 $\Lambda_c^+\rightarrow\bar{K}^0\eta p$ processes,
 in spite of this
 channel
 having
 a large coupling to the $N^*(1650)$.
 This is because the $K\Sigma$ pair production is suppressed in the
 primary production from the $\Lambda_c$ decay.
\end{abstract}

\maketitle
\section{Introduction}
The nature of the $N^*$(1535) ($J^P = 1/2^-$) remains to be well
understood \cite{Klempt:2007cp,Crede:2013sze}. Its properties have been
studied within the context of the constituent quark model
\cite{Capstick:2000qj,Patrignani:2016xqp} where
the mass of the lowest excitation of the nucleon with a negative parity
is
found
smaller than its positive-parity counterpart,
contrary to what is observed in experiment,
namely, the $N^*(1535)$ and $N^*(1440)$ resonances.
This is known as the mass reverse problem.
Also it seems to be difficult to explain the fact that the
$N^*$(1535) could couple to channels with strangeness, such as $\eta N$
and $K \Lambda$ \cite{Patrignani:2016xqp,Liu:2005pm},
within the formalism of the quark model with a simple $qqq$
configuration where the $\bar{s}s$ component is not
contained in the $N^*(1535)$ resonance.
Studies, such
as the ones found in
Refs.~\cite{Glozman:1997ag,Bijker:1994yr,Helminen:2000jb,An:2008xk,Ferretti:2015ada},
attempt to solve some difficulties in the description of the $N^*(1535)$
properties with some extension of the conventional quark model, and
the possible role of the $N^*(1535)$ resonance in some reactions is
explored in
Refs.~\cite{Li:1996wj,Doring:2005bx,Xie:2007qt,Cao:2008st,Geng:2008cv,Doring:2008sv,Cao:2009ea,Debastiani:2017dlz}.

On the other hand, by using the chiral Lagrangians within the framework of the
unitary coupled channels approach,
some previously unexplained baryonic resonances could be understood as
meson-baryon molecular states.
{A well-known example of this are the studies of the $\Lambda$(1405)
that were carried out in
Refs.~\cite{Kaiser:1996js,Kaiser:1995eg,Oset:1997it,Jido:2003cb,Oller:2000fj,GarciaRecio:2002td,Hyodo:2007jq,Hyodo:2007np,Hyodo:2011ur,Kamiya:2016jqc}.
In the same way, the $N^*(1535)$ resonance is studied including the $\eta N$,
$\pi N$, $K\Lambda$ and $K\Sigma$ channels.}
The mass and width of the $N^*$(1535)
could be obtained by calculating the position of the poles of the $T$
matrix on the second (unphysical) Riemann sheet
\cite{Nieves:2001wt,Inoue:2001ip,Bruns:2010sv,Gamermann:2011mq,Khemchandani:2013nma,Garzon:2014ida},
and were found to be in good agreement with experiment. Using this
formalism, the $N^*$(1535) was also found to couple strongly to $\eta
N$, $K \Sigma$ and $K \Lambda$, as well as less strongly to $\pi N$. In
Refs.~\cite{Kaiser:1996js,Kaiser:1995cy,Inoue:2001ip} in particular, where the
$N^*$(1535) was dynamically generated through pseudoscalar
meson--baryon ($PB$) interactions.
The loop functions were renormalized
using the cutoff (in Refs.~\cite{Kaiser:1996js,Kaiser:1995cy}) and dimensional (in
Ref.~\cite{Inoue:2001ip}) regularization schemes,
and
the
cutoffs/subtraction constants were required to have different values
for each of the coupled channels in order to get a good agreement with
experiment. This is quite different from the case of the
$\Lambda$(1405), where only a single global cutoff was needed
\cite{Oset:1997it}.
In the case of the dimensional regularization \cite{Inoue:2001ip},
the values of the subtraction constants are different from the
``natural'' size which is related to the
mass of the first resonance ({the} $\rho$ meson in this case)
\cite{Oller:2000fj}.
On the other hand{, from the consideration of the Castillejo-Dalitz-Dyson pole contribution,
the study of Ref.~\cite{Hyodo:2008xr} suggests that {some}
contribution other than the meson--baryon component would also be important
for the $N^*(1535)$.}

In the vector meson-baryon system, the $N^*(1650)$ was firstly obtained as
a degenerate state of $J^P=1/2^-$ and $3/2^-$ in the study of
the vector octet-baryon octet system with the chiral unitary approach \cite{Oset:2009vf}.
The $J^P=3/2^-$ {case} was studied in Ref.~\cite{Garzon:2013pad} with the
$\rho N$($s$ wave), $\pi\Delta$($s$ wave), $\pi N$($d$ wave) and
$\pi\Delta$($d$ wave) channels, and there
a pole
was found
which can be associated with the $N^*(1700)$ resonance,
having a sizable
coupling to
$\rho N$.
The mixing effects
of
$PB$ channels with vector meson--baryon ($VB$)
channels
with $J^P=1/2^-$
were explored in
Refs.~\cite{Garzon:2012np,Khemchandani:2011mf,Khemchandani:2012ur,Khemchandani:2013nma}
and they were found to be quite significant. In
Ref.~\cite{Garzon:2014ida} the possibility that the missing component in
Refs.~\cite{Kaiser:1995cy,Inoue:2001ip} corresponds to $VB$ channels was
explored by introducing the $\rho N$($s$ wave) and $\pi \Delta$($d$
wave) states
in
the model
of
Ref.~\cite{Inoue:2001ip} using the local hidden gauge
formalism. Doing this, both the $N^*$(1535) and $N^*$(1650) ($J^P =
1/2^-$) resonances were dynamically generated, and
the
masses and
widths obtained were very close to their experimental values. Also the
subtraction constants used
in that study, although still different for each
channel, were now very close to a
``natural'' value. A similar work
to this was done in Ref.~\cite{Khemchandani:2013nma}. The two resonances
were also generated in Refs.~\cite{Nieves:2001wt,Bruns:2010sv} using
only $PB$ channels with an off-shell approach that is equivalent to
considering different subtraction constants from those in
Ref.~\cite{Garzon:2014ida}.

Nonleptonic weak decays have been widely explored with the objective of
studying and testing the properties of baryonic resonances
\cite{Crede:2008vw,Chen:2016qju,Miyahara:2015cja,Hyodo:2011js,Xie:2016evi,Oset:2016lyh,Xie:2017erh},
thus allowing for a way to distinguish between the different models used
to generate them. For example, in Ref.~\cite{Hyodo:2011js} the decay
$\Lambda_c^+ \rightarrow \pi^+ \pi \Sigma$ was studied in order to get
the $\pi \Sigma$ scattering lengths. In Ref.~\cite{Miyahara:2015cja} the
$\Lambda_c^+ \rightarrow \pi^+ MB$ decay, with the $M$ a meson and $B$ a
baryon, for $MB=\pi\Sigma,\bar{K}N,$ and $\eta\Lambda$ was studied to
better understand the $\Lambda$(1405) and $\Lambda$(1670) properties,
and in Ref.~\cite{Xie:2016evi} the $\Lambda_c^+ \rightarrow \eta \pi^+
\Lambda$ was used to
{investigate}
the $a_0$(980) and
$\Lambda$(1670) resonances. With this in mind, in
Ref.~\cite{Xie:2017erh} the $\Lambda_c^+ \rightarrow \bar{K}^0 \eta p$
decay was used to study the nature of the $N^*$(1535) by comparing
different models, including the one in Ref.~\cite{Inoue:2001ip}. In that
study, only $PB$ channels were considered in this process, which
corresponds to ignoring the influence that the $VB$ channels can have in
the nonleptonic decay
through
a large coupling of the $N^*(1535)$ to the
$\rho N$ channel,
as
found in Ref.~\cite{Garzon:2014ida}.
Indeed, the effect of
the $VB$ channel can be quite large in some reactions {as was shown in}
Refs.~\cite{Liang:2016ydj,Pavao:2017cpt}.

In this paper we extend the calculations done in Ref.~\cite{Xie:2017erh}
to take into account the $VB$ channels, and
the effects of the
$N^*$(1650) resonance, using the model developed in
Ref.~\cite{Garzon:2014ida}. Using this we calculate the mass
distribution of $\eta N$ in the $\Lambda_c^+ \rightarrow \bar{K}^0 \eta
p$ decay and the mass distribution of $\pi N$ and $K\Sigma$ in the
$\Lambda_c^+ \rightarrow \bar{K}^0 \pi N$,
$\Lambda_c^+\rightarrow\bar{K}^0K\Sigma$ decays.
In this way we hope to shed some light on the nature of the $N^*$(1535) as well as the $N^*$(1650).

The paper is organized as follows.
The theoretical framework of this study, the weak process of
$\Lambda_c^+\rightarrow \bar{K}^0MB$ and the meson-baryon scattering
amplitude, is given in Sec.~\ref{sec_formalism}.
Section~\ref{sec_results} is devoted to the results,
the mass distribution of the $\Lambda_c^+\rightarrow \bar{K}^0MB$
$[MB=\pi N(I=1/2),\eta p$ and $K\Sigma(I=1/2)]$.
A summary of this work is given in Sec.~\ref{sec_summary}.

\section{Formalism}
\label{sec_formalism}
\begin{figure}[t]
 \centering
 \includegraphics[width=8cm]{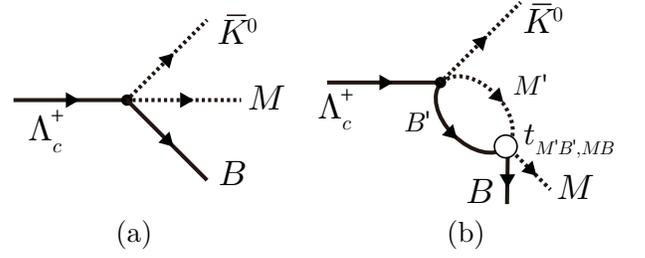}
 \caption{The diagrams for the $\Lambda_c^+\rightarrow \bar{K}^0MB$
 decay.}
 \label{fig_tree-loop}
\end{figure}

The diagrams for the $\Lambda_c^+$ decay into $\bar{K}^0MB$ which we
take into account in this study
are depicted in Fig.~\ref{fig_tree-loop}.
The primary $\bar{K}^0MB$ production in the
$\Lambda_c^+$ decay occurs in the weak process
and
it is
followed by the rescattering of the meson-baryon pair $MB$
where, as studied in Ref.~\cite{Garzon:2014ida},
the resonances $N^*(1535)$ and $N^*(1650)$ are generated through the
dynamics of hadrons.

\begin{figure}[t]
 \centering
 \includegraphics[width=8cm]{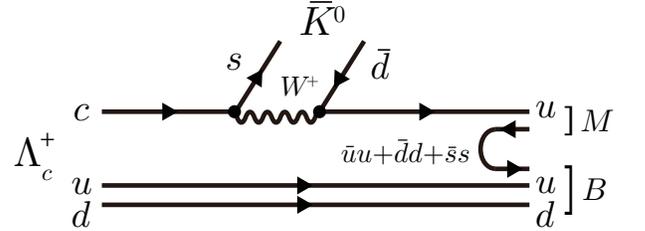}
 \caption{The quark-level diagram for the $\Lambda_c^+\rightarrow
 \bar{K}^0MB$ process.}
 \label{fig_production}
\end{figure}

First
we discuss the primary vertex of the $\Lambda_c^+$
decay into $\bar{K}^0MB$.
In this process, we use the
same
approach
as done in
Ref.~\cite{Xie:2017erh}, but now we have an additional $\rho
N$ channel.
We consider the diagram shown in Fig.~\ref{fig_production} for the weak
transition and the hadronization
at
the quark level.
The reaction can occur with the intermediate $W^+$ exchange
with the Cabibbo-allowed coupling of $W^+$ to $cs$ and $\bar{d}u$
\cite{Chau:1982da},
{with a sequential} pair creation of the light quark from
the vacuum.
The $\bar{d}s$ pair forms the $\bar{K}^0$, and the remaining $uud$
quarks with a $\bar{q}q$ from the vacuum hadronize into the
meson-baryon pair.
In this approach, the $ud$ pair in $\Lambda_c^+$ with the spin $S=0$ and
isospin $I=0$
acts
as a spectator.
Then,
at the quark level we can write the final state as
\begin{align}
 u(\bar{u}u+\bar{d}d+\bar{s}s)\frac{1}{\sqrt{2}}(ud-du)=\sum_iM_{1i}q_i\frac{1}{\sqrt{2}}(ud-du)
\end{align}
where $M_{ij}=q_i\bar{q}_j$ $(q_1=u,q_2=d,q_3=s)$.
{The matrix $M$
at
the quark level can be related with that
at
the hadronic
level based on the flavor symmetry.}
Then,
the matrix $M$
for the pseudoscalar meson
is
given
by
\cite{Pavao:2017cpt}
\begin{align}
 M=
 \begin{pmatrix}
  \frac{\pi^0}{\sqrt{2}}+\frac{\eta}{\sqrt{3}}+\frac{\eta'}{\sqrt{6}}&\pi^+&K^+\\
  \pi^-&-\frac{\pi^0}{\sqrt{2}}+\frac{\eta}{\sqrt{3}}+\frac{\eta'}{\sqrt{6}}&
  K^0\\
  K^-&\bar{K}^0 &-\frac{\eta}{\sqrt{3}}+\sqrt{\frac{2}{3}}\eta' 
 \end{pmatrix}
\end{align}
which contains the $\eta$, $\eta'$ mixing of Ref.~\cite{bramon},
and
we obtain
\begin{align}
 &\sum_iM_{1i}q_i\frac{1}{\sqrt{2}}(ud-du)=\left(\frac{\pi^0}{\sqrt{2}}+\frac{\eta}{\sqrt{3}}\right)u\frac{1}{\sqrt{2}}(ud-du)\notag\\
 &+\pi^+d\frac{1}{\sqrt{2}}(ud-du)+K^+s\frac{1}{\sqrt{2}}(ud-du).
\end{align}
{Referring to Ref.~\cite{Miyahara:2016yyh} for
the quark representation of the baryons
(see also footnote~$1$ in Ref.~\cite{Pavao:2017cpt}),}
\begin{align}
 p=&\frac{u\left(ud-du\right)}{\sqrt{2}},\\
 n=&\frac{d\left(ud-du\right)}{\sqrt{2}},\\ 
 \Lambda=&\frac{u\left(ds-sd\right)+d\left(su-us\right)-2s\left(ud-du\right)}{2\sqrt{3}},
\end{align}
we can write the final state of the
pseudoscalar meson and baryon {$\left|PB\right>$,
apart from
the $\bar{K}^0$ meson,
 as}
\begin{align}
 \left|PB\right>=&\frac{1}{\sqrt{2}}\left|\pi^0p\right>+\frac{1}{\sqrt{3}}\left|\eta
 p\right>+\left|\pi^+n\right>-\sqrt{\frac{2}{3}}\left|K^+\Lambda\right>\notag\\
 =&-\sqrt{\frac{3}{2}}\left|\pi
 N(I=1/2)\right>+\frac{1}{\sqrt{3}}\left|\eta p\right>-\sqrt{\frac{2}{3}}\left|K^+\Lambda\right>,\label{eq_pb_chan}
\end{align}
where the $\pi N$ channel is written in terms of the isospin basis
($\left|\pi^+\right>=-\left|I=1,I_z=1\right>$ in this convention).
Here, we have omitted the $\eta'p$ channel because the threshold is far
above the energy of the $N^*(1535)$ and $N^*(1650)$ that we focus on in this study.
In the same way, 
{replacing the matrix $M$ with the matrix $V$ for the vector mesons~\cite{Oset:2009vf}}
\begin{align}
 V=
 \begin{pmatrix}
  \frac{\rho^0}{\sqrt{2}}+\frac{\omega}{\sqrt{2}}&\rho^+&K^{*+}\\
  \rho^-&-\frac{\rho^0}{\sqrt{2}}+\frac{\omega}{\sqrt{2}} &K^{*0} \\
  K^{*-}&\bar{K}^{*0} &\phi 
 \end{pmatrix},
\end{align}
where the ideal mixing of the isospin-singlet mesons is assumed,
we can obtain the final state with a vector meson $\left|VB\right>$ as
\begin{align}
 \left|VB\right>= -\sqrt{\frac{3}{2}}\left|\rho N(I=1/2)\right>.\label{eq_vb_chan}
\end{align}
Here, the irrelevant channels containing the $\omega$, $\phi$, $K^*$ and
$\bar{K}^*$ mesons
are omitted and the phase convention $\left|\rho^+\right>=-\left|I=1,I_z=1\right>$
should be understood.

Combining these two cases {in} Eqs.~(\ref{eq_pb_chan}) and
(\ref{eq_vb_chan}), we can write the hadronic final state except for the
$\bar{K}^0$ meson $\left|MB\right>$ as
\begin{align}
 \left|MB\right>=&-\sqrt{\frac{3}{2}}\left|\pi
 N(I=1/2)\right>+\frac{1}{\sqrt{3}}\left|\eta
 p\right>-\sqrt{\frac{2}{3}}\left|K^+\Lambda\right>\notag\\
 &-\sqrt{\frac{3}{2}}\left|\rho N(I=1/2)\right>\notag\\
 \equiv&\sum_{MB}h_{MB}\left|MB\right>,\label{eq_final_MB}
\end{align}
{where} the coefficient of each channel $h_{MB}$ stands for the relative production
weight from the $\Lambda_c^+$ and is summarized in
Table~\ref{tab_coefficient}.
\begin{table}[t]
 \centering
 \begin{tabular}[t]{c|cccc}
  &$\pi N(I=1/2)$ &$\eta N$ &$K\Lambda$ &$\rho N(I=1/2)$ \\\hline
  $h_{MB}$& $-\sqrt{\frac{3}{2}}$ &$\frac{1}{\sqrt{3}}$ &$-\sqrt{\frac{2}{3}}$ &$-\sqrt{\frac{3}{2}}$ \\
  $f_{MB}$& $\frac{1}{4\pi}\frac{1}{2}$
      &$\frac{1}{4\pi}\frac{1}{2}$
	  &$\frac{1}{4\pi}\frac{1}{2}$
	      &$\frac{1}{4\pi}\frac{1}{2\sqrt{3}}$ \\
 \end{tabular}
 \caption{The table for the coefficients $h_{MB}$ and $f_{MB}$ in Eq.~(\ref{eq_production_amp}).}
 \label{tab_coefficient}
\end{table}
The weight of the $\rho N$ channel in
Eq.~(\ref{eq_final_MB}) is only due to flavor.
{In addition a different spin structure of the pseudoscalar and vector meson leads to
a different factor for the production weight in the decay process, as was studied  
in Refs.~\cite{Liang:2016ydj,Pavao:2017cpt} based on the $^3P_0$ model for the hadronization.}
Now, we only need to see the $J=1/2$ case
because the resonances $N^*(1535)$ and $N^*(1650)$ have $J^P=1/2^-$.
Because the $q\bar{q}$ should have $J^P=0^+$ which are the same quantum numbers as those
of the vacuum, the total angular momentum after the hadronization should
come from that of the $u$ quark from the weak vertex that is denoted by $\left|J,M;u\right>$.
According to the $^3P_0$ model \cite{Micu:1968mk,LeYaouanc:1972vsx,Close:1979bt},
the angular momentum $L$ should be $L=1$ for parity conservation,
and at the same time the spin $S$ should be $S=1$ to have $J=0$ by
addition with $L=1$. 
This is written as $\left|0,0;\bar{q}q\right>_{^3P_0}$.
The $ud$ pair
in the $\Lambda_c^+$, or equivalently in
the final state baryon, has spin $J=0$ and isospin $I=0$ that is
written as $\left|0,0;ud\right>_{\rm spectator}$.
Following {the} works of Refs.~\cite{Liang:2016ydj,Pavao:2017cpt},
writing the relative angular momentum between the produced
$u$ quark from the weak vertex and $\bar{q}$ of the $\bar{q}q$ from the
vacuum in the final state as $j$, we can rewrite the spin structure of the
system as
\begin{align}
 &\left|J,M;u\right>\left|0,0;\bar{q}q\right>_{^3P_0}\left|0,0;ud\right>_{\rm
 spectator}\notag\\
 &\hspace{2cm}=\sum_{j}\mathcal{C}(j,J)\left|J,M,j\right>.
\end{align}
Now,  the $j=0$ and $1$ cases correspond to the pseudoscalar and vector
meson production, respectively.
{Then, since we are only interested in the $J=1/2$ case, we can write}
\begin{align}
 &\left|\frac{1}{2}, \pm \frac{1}{2};u\right>\left|0,0;\bar{q}q\right>_{^3P_0}\left|0,0;ud\right>_{\rm
 spectator}\notag\\
 &\hspace{2cm}=\sum_{MB}f_{MB}\left|\frac{1}{2},\pm \frac{1}{2};MB\right>,\label{eq_spin_final}
\end{align}
where the factor $f_{MB}$ is
$\frac{1}{4\pi}\frac{1}{2}$ and $\frac{1}{4\pi}\frac{1}{2\sqrt{3}}$ for the
cases with $M$ the pseudoscalar meson and the vector meson,
respectively, and we show it in Table~\ref{tab_coefficient}.

Then, we can write the decay amplitude of the tree-level diagram given in
Fig.~\ref{fig_tree-loop}(a) as
\begin{align}
 t_{\Lambda_c\rightarrow \bar{K}^0MB}=&V_Ph_{MB}f_{MB},\label{eq_production_amp}
\end{align}
where $V_P$ is a common constant for the strength of the production and
the coefficients $h_{MB}$ and $f_{MB}$ are the factors originating from the flavor and
spin structures given in Eqs.~(\ref{eq_final_MB}) and
(\ref{eq_spin_final})
(see Table~\ref{tab_coefficient}).
In this study, we omit the possible energy dependence of the amplitude
because the reaction proceeds in $s$ wave and, as we will see later,
{only} a small energy range around the $N^*(1535)$ and $N^*(1650)$ resonances is
of our interest.

In this approach, the $\pi\Delta$ and $K\Sigma$ productions are suppressed
because the $ud$ pair in $\Lambda_c^+$, which has
spin $S=0$ and
isospin $I=0$, is
a spectator, $i.e.$, the spin and isospin
structure of the $ud$ pair is not changed throughout the hadronization process.
While there are other possibilities for the creation of the quark
pair which enable us to have the $K\Sigma$ or $\pi\Delta$ production,
the study of Ref.~\cite{Wang:2015pcn} suggests that in the case of
$\Lambda_b^0\rightarrow J/\psi\pi^-p$, which has the same topology
as
the diagram of the weak process
studied here,
the spectator treatment gives a
good description for the experimental data of Ref.~\cite{Aaij:2014zoa}.
Then, we expect {that} this treatment {also works well in the present case.}

For the meson-baryon amplitude $t_{MB,M'B'}$ in
Fig.~\ref{fig_tree-loop}(b) which is responsible for the rescattering
after the hadronization,
we follow the study of Ref.~\cite{Garzon:2014ida}.
In the study, the meson-baryon amplitude
was
evaluated by
using
the
chiral unitary approach with the $\pi N$, $\eta N$, $K\Lambda$,
$K\Sigma$, $\rho N$, and $\pi\Delta$($d$ wave) channels,
and it was found that the $N^*(1535)$ and $N^*(1650)$ resonances are
dynamically generated.
The interaction kernel of $PB$ to $PB$ and $VB$ to $VB$ is given by the
leading order of the chiral Lagrangian, or equivalently the vector meson
exchange \cite{Inoue:2001ip,Oset:2009vf}, and the transition of $PB$ to
$VB$ is taken into account through the one pion exchange {and the
Kroll-Ruderman term}
\cite{Garzon:2012np,Garzon:2013pad}\footnote{In practice to
obtain the
same
result to Ref.~\cite{Garzon:2014ida},
we add the {contact {and Born} terms} to the diagonal $\rho N$ channel of the interaction kernel
as in Ref.~\cite{Khemchandani:2012ur},
and the {energy} transfer in the one pion exchange diagram is omitted in
this calculation.}.
Then, writing the interaction kernel as $v$, the meson-baryon amplitude
$t_{MB,M'B'}$ is given by
\begin{align}
 t_{MB,M'B'}=\left[\left(1-vG\right)^{-1}v\right]_{MB,M'B'},
\end{align}
where $G$ is the meson-baryon loop function evaluated with
dimensional regularization.
The analytic form of the loop function of the $MB$ channel,
$G_{MB}(\sqrt{s},m_M,M_B)$, is given by
\begin{align}
 &G_{MB}(\sqrt{s},m_M,M_B)=\frac{2M_B}{16\pi^2}\{a_{MB}(\mu)+\ln\frac{M_B^2}{\mu^2}\notag\\
 &\hspace{5mm}+\frac{m_M^2-M_B^2+s}{2s}\ln\frac{m_M^2}{M_B^2}\notag\\
 &\hspace{5mm}+\frac{q_{MB}}{\sqrt{s}}\left[\ln(s-M_B^2+m_M^2+2q_{MB}\sqrt{s})\right.\notag\\
 &\hspace{1.5cm}+\ln(s+M_B^2-m_M^2+2q_{MB}\sqrt{s})\notag\\
 &\hspace{1.5cm}-\ln(-s+M_B^2-m_M^2+2q_{MB}\sqrt{s})\notag\\
 &\hspace{1.5cm}\left.-\ln(-s-M_B^2+m_M^2+2q_{MB}\sqrt{s})\right]\},\label{eq_loop_anal}
\end{align}
with $\mu$ the regularization scale, $m_M$ and $M_B$ the mass of the meson
and baryon, respectively, and $q_{MB}$ the meson momentum in the
meson-baryon center-of-mass (CM) frame $q_{MB}=\lambda^{1/2}(s,m_M^2,M_B^2)/2\sqrt{s}$ where
$\lambda(x,y,z)=x^2+y^2+z^2-2xy-2yz-2zx$.

Finally, the decay amplitude of the $\Lambda_c^+\rightarrow \bar{K}^0MB$
process from the diagrams in Fig.~\ref{fig_tree-loop}(a) and (b) is
given by
\begin{align}
 &t_{\Lambda_c^+\rightarrow \bar{K}^0MB}=V_Ph_{MB}f_{MB} \notag\\
 +&\sum_{M'B'}V_Ph_{M'B'}f_{M'B'}G_{M'B'}(M_{M'B'})t_{M'B',MB}(M_{MB}),\label{eq_amp_final}
\end{align}
where $M_{MB}$ denotes the invariant mass of the meson $M$ and baryon
$B$ (now $M_{MB}=M_{M'B'}$).
Regarding the meson-baryon loop function
$G_{MB}$
following
the tree-level amplitude for $\Lambda_c^+\rightarrow\bar{K}^0M'B'$
and
before
$t_{M'B',MB}$ in Fig.~\ref{fig_tree-loop}(b),
we use the same subtraction constants as those in the meson-baryon amplitude
$t_{MB,M'B'}$ given in Ref.~\cite{Garzon:2014ida}.
In the same way as
done
in
Refs.~\cite{Garzon:2014ida,Garzon:2013pad}
we use the $\rho N$ loop
function $\tilde{G}_{\rho N}$ which is obtained by smearing the loop
function
$G_{\rho N}(\sqrt{s},m_\rho,M_N)$ given by Eq.~(\ref{eq_loop_anal}) with
the $\rho$-meson spectral
function to take account of the width of the $\rho$ meson\footnote{We note that the real part of the $\rho N$ loop
function becomes positive below the $\rho N$ threshold with the
subtraction constant in Ref.~\cite{Garzon:2014ida}.},
\begin{align}
 &\tilde{G}_{\rho
 N}(\sqrt{s})=\frac{1}{N}\int_{m_\rho-2\Gamma_\rho}^{m_\rho+2\Gamma_\rho}2\tilde{m}d\tilde{m}\left(-\frac{1}{\pi}\right)\notag\\
 &\hspace{5mm}\cdot\im\left[\frac{1}{\tilde{m}^2-m_\rho^2+i\tilde{m}\Gamma_\rho(\tilde{m})}\right]G_{\rho N}(\sqrt{s},\tilde{m},M_N)
\end{align}
with
\begin{align}
 \Gamma_\rho(\tilde{m})=&\Gamma_\rho\frac{|\vec{q}\,|^3}{|\vec{q}\,|^3_{\rm
 on}}\theta(\tilde{m}-2m_\pi), \\
 |\vec{q}\,|=&\frac{\lambda^{1/2}(\tilde{m}^2,m_\pi^2,m_\pi^2)}{2\tilde{m}},\\
 |\vec{q}\,|_{\rm
 on}=&\frac{\lambda^{1/2}(m_\rho^2,m_\pi^2,m_\pi^2)}{2m_\rho},\\
 N=&\int_{m_\rho-2\Gamma_\rho}^{m_\rho+2\Gamma_\rho}2\tilde{m}d\tilde{m}\left(-\frac{1}{\pi}\right)\notag\\
 &\cdot\im\left[\frac{1}{\tilde{m}^2-m_\rho^2+i\tilde{m}\Gamma_\rho(\tilde{m})}\right].
\end{align}

Here, we note that the $K\Sigma$ and $\pi\Delta$ channels are not
included in the sum of $M'B'$ in Eq.~(\ref{eq_amp_final}) because there
is no direct production from $\Lambda_c^+$ in our approach
in Eq.~(\ref{eq_final_MB}), while these channels
appear in
the meson-baryon amplitude $t_{MB,M'B'}$.

With an appropriate phase-space factor, the mass distribution
$d\Gamma_{\Lambda_c^+\rightarrow\bar{K}^0MB}/dM_{MB}$ as a function of
$M_{MB}$ is given by
\begin{align}
 \frac{d\Gamma_{\Lambda_c^+\rightarrow\bar{K}^0MB}}{dM_{MB}}=\frac{1}{(2\pi)^3}\frac{M_B}{\mlamc}|\vec{p}_{\bar{K}^0}||\vec{\tilde{p}}_{M}||t_{\Lambda_c^+\rightarrow
 \bar{K}^0MB}|^2,
\end{align}
where
$p_{\bar{K}^0}$ and $\tilde{p}_{M}$ are the momentum of $\bar{K}^0$ in the $\Lambda_c^+$ rest frame
and that of the meson $M$ in the $MB$ CM frame, respectively, with
\begin{align}
 |\vec{p}_{\bar{K}^0}|=&\frac{\lambda^{1/2}(\mlamc^2,m_{\bar{K}^0}^2,M_{MB}^2)}{2\mlamc}, \\
 |\vec{\tilde{p}}_{M}|=&\frac{\lambda^{1/2}(M_{MB}^2,m_M^2,M_B^2)}{2M_{MB}}.
\end{align}

Here, we give a comment on the possible modification of the mass
distribution by the rescattering of $\bar{K}^0$ with the meson $M$ or
baryon $B$ in the final state, which are not taken into account in this
study.
The $\bar{K}^0p$ in the
$\Lambda_c^+\rightarrow\bar{K}^0\eta p$ decay can couple to some
$\Sigma^*$ resonances, but as pointed out in Ref.~\cite{Xie:2017erh},
these resonances would not give a large modification
to the mass distribution because of the small overlap with the phase
space and the $p$-wave coupling of the $\Sigma^*$ to the $\bar{K}^0p$
channel.
Another possibility is the coupling of $K\bar{K}$ with the $a_0(980)$
or $f_0(980)$ states in the $\Lambda_c^+\rightarrow\bar{K}^0K\Lambda$ or
$\bar{K}^0K\Sigma$ decays.
In this case, the invariant mass of the $\bar{K}^0K$ pair
spreads up in a range of invariant masses above
1050~MeV,
and
then, the overlap of the $a_0(980)$ and $f_0(980)$ resonances with the
$\Lambda_c^+\rightarrow\bar{K}^0K\Lambda$ or $\bar{K}^0K\Sigma$ phase
space is small.
Though some $\Lambda^*$ resonances can also contribute in the
$\Lambda_c^+\rightarrow\bar{K}^0\pi N$ process through the $\bar{K}^0N$
rescattering, it does not matter in our case because now we are
interested in the mass distribution as a function of $M_{\pi N}$, not
$M_{\bar{K}^0N}$, where the $\Lambda^*$ distributes its strength.
Then, a resonance such as $\Lambda(1800)$ \cite{Patrignani:2016xqp} which can have a
certain overlap with the phase space in
$d^2\Gamma_{\Lambda_c^+\rightarrow\bar{K}^0\pi N}/dM_{\pi
N}dM_{\bar{K}^0N}$ is integrated over in $M_{\bar{K}N}$, and gives just a
broad background in the $M_{\pi N}$ mass distribution.

\section{Results}
\label{sec_results}
\begin{figure}[t]
 \centering
 \includegraphics[width=7.5cm]{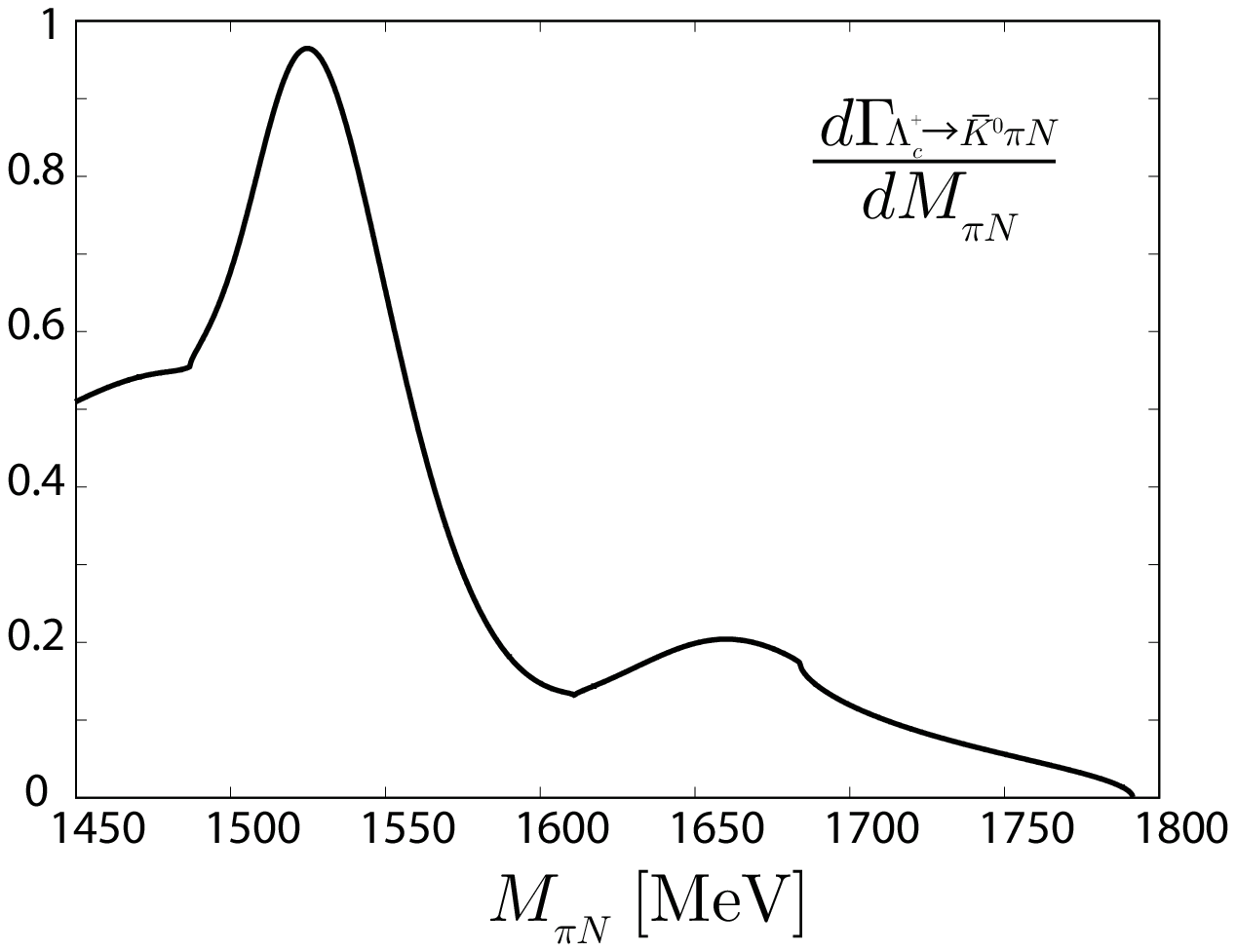}
 \includegraphics[width=7.5cm]{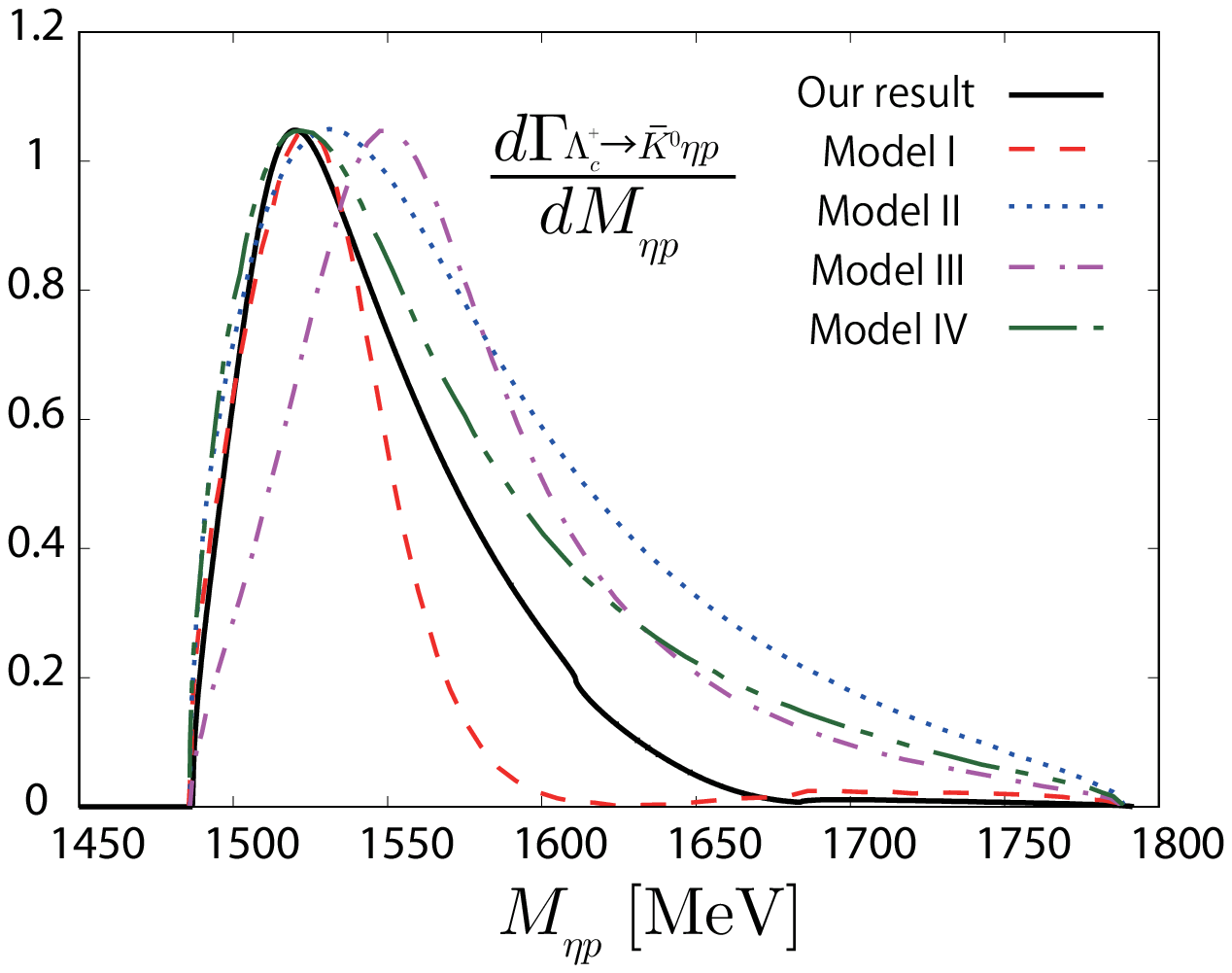}
 \includegraphics[width=7.5cm]{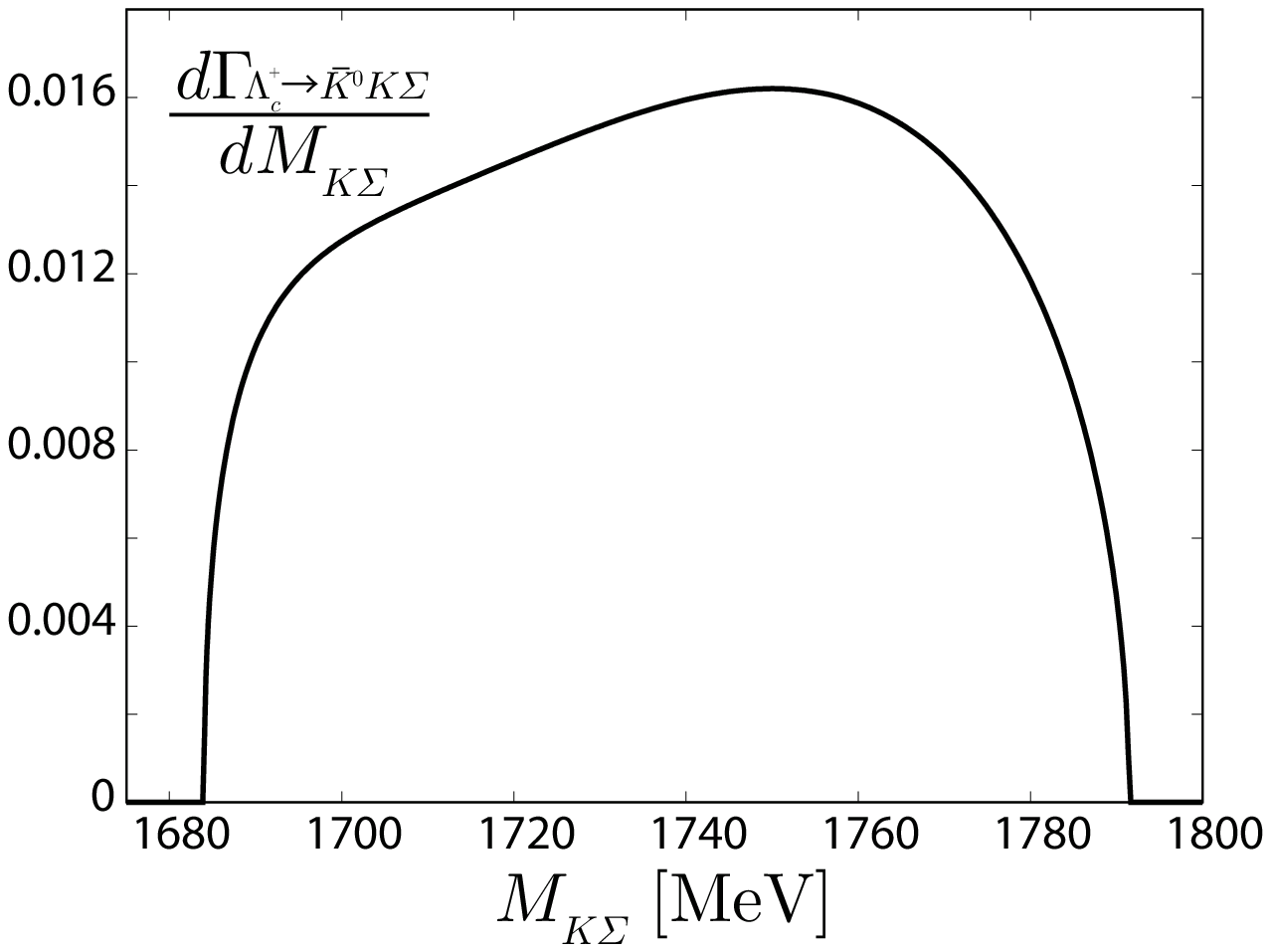}
 \caption{The mass distributions for $\Lambda_c^+$ decay into
 $\bar{K}^0\pi N$ with $I=1/2$ (top), $\bar{K}^0\eta p$ (middle), and
 $\bar{K}^0K\Sigma$ with $I=1/2$ (bottom) as functions of $M_{MB}$.
 {In the middle figure, the lines other than the solid one are the results given in
 Ref.~\cite{Xie:2017erh} with the height scaled to agree with the
 result of this study.}}
 \label{fig_massdist_1}
\end{figure}

The mass distributions
$d\Gamma_{\Lambda_c^+\rightarrow\bar{K}^0MB}/dM_{MB}$ with $MB=\pi
N(I=1/2),\eta p$, and $K\Sigma(I=1/2)$ as functions of $M_{MB}$ are given in
Fig.~\ref{fig_massdist_1}.
In these figures, we show the results with $V_P=1~\mev^{-1}$ because of
our lack of the knowledge to fix the value of $V_P$.
{This is not a problem since we only want to focus on the
behavior of the mass distribution.}

\begin{table}[t]
 \centering
 \begin{tabular}[t]{c||c|c|c|c|c|c}
  &$\pi N$ &$\eta N$ &$K\Lambda$ &$K\Sigma$&$\rho N$ &$\pi \Delta$ \\\hline\hline
  $N^*(1535)$& $25.2$ &$42.2$ &$40.7$ &$3.2$ &$17.9$ &$8.8$ \\\hline
  $N^*(1650)$& $36.6$ &$34.0$ &$20.3$ &$31.6$ &$8.1$ &$9.0$ \\
 \end{tabular}
 \caption{The absolute values of $g_{N^*,M'B'}G_{M'B'}$ (in MeV)
 {at the resonance pole,}
 taken from Ref.~\cite{Garzon:2014ida}.}
 \label{tab_gG}
\end{table}

\begin{figure}[t]
 \centering
 \includegraphics[width=5cm]{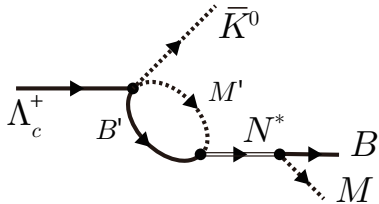}
 \caption{The diagram from the resonance for the
 $\Lambda_c^+\rightarrow\bar{K}^0MB$.}
 \label{fig_resonance}
\end{figure}

For the $\pi N$ mass distribution of the $\Lambda_c^+\rightarrow\bar{K}^0\pi N$
decay, we can see two peaks;
the peak located {in the lower} energy, which is associated with the $N^*(1535)$
resonance, has larger strength than the one in the higher energy which
comes from the $N^*(1650)$.
On the other hand in the
scattering
amplitude of the diagonal $\pi N$ channel in
Ref.~\cite{Garzon:2014ida},
the magnitude of the higher peak is
larger than that of the lower peak.
We can understand this difference from the coupling of the resonances
with the meson-baryon states given in Ref.~\cite{Garzon:2014ida}.
Indeed, $g_{N^*(1535),\pi N}=1.03+i0.21$ versus $g_{N^*(1650),\pi
N}=1.37+i0.54$.
Then, the Breit-Wigner amplitude $g_{R,\pi
N}^2/(\sqrt{s}-M_{R}+i\Gamma_R/2)$ has larger strength in the case of
the $N^*(1650)$.
On the other hand,
if we write the meson-baryon amplitude with the Breit-Wigner amplitude
(see Fig.~\ref{fig_resonance} for the diagram), the $\Lambda_c^+\rightarrow
\bar{K}^0MB$ amplitude $T_{BW}$ is given by
\begin{align}
 T_{BW}=&\sum_{N^*}\sum_{M'B'}V_Ph_{M'B'}f_{M'B'}G_{M'B'}(M_{M'B'})\notag\\
 &\cdot\frac{g_{N^*,M'B'}g_{N^*,MB}}{M_{MB}-M_{N^*}+i\Gamma_{N^*}/2},
\end{align}
where the sum of $N^*$ runs over $N^*(1535)$ and $N^*(1650)$.
Then, the difference of the intermediate states appears in the
combination of $g_{N^*,M'B'}G_{M'B'}$
around
the resonance peak.
We compare the absolute values of $g_{N^*,M'B'}G_{M'B'}$, {given here in Table~\ref{tab_gG}}, to get a rough understanding.
The value of $g_{N^*,MB}G_{MB}$ for the $\eta N$ and $K\Lambda$ channels
is larger for $N^*(1535)$ than $N^*(1650)$ while the magnitude of the
coupling of the $\pi N$ channel to $N^*(1650)$ is
larger than the coupling to $N^*(1535)$.
Furthermore, in the primary vertex the $K\Sigma$ channel which has a larger
coupling to $N^*(1650)$ than $N^*(1535)$ is not produced.
As the result, the peak of the $N^*(1535)$ resonance is larger than
that of the $N^*(1650)$ in the mass distribution of $\Lambda_c^+$ decay into
$\bar{K}^0\pi N$.

{At the middle of Fig.~\ref{fig_massdist_1}, we show the $\eta p$
invariant mass distribution in the $\Lambda_c^+\rightarrow\bar{K}^0\eta
p$ process with the result of Ref.~\cite{Xie:2017erh} for
comparison. 
In this case,} we
can see only a single peak.
{Compared with the mass distribution of Model~I in
Ref.~\cite{Xie:2017erh}, the mass distribution has a larger width.
This would
be attributed to the effect of the
$N^*(1650)$, analogously to the amplitude of the $\pi N$ to $\eta N$
reaction in Ref.~\cite{Garzon:2014ida} where a single peak is
observed in the cross section and its larger width than
in
Ref.~\cite{Inoue:2001ip} is ascribed to the $N^*(1650)$.
On the other hand,}
the contribution from the $N^*(1650)$ is more suppressed
than that in the $\Lambda_c^*\rightarrow\bar{K}^0\pi N$ process because
{of the stronger coupling of the $\eta N$ channel to the $N^*(1535)$
than $N^*(1650)$
In addition,
the absence of
the $K\Sigma$ channel in the initial production process (see
Eq.~(\ref{eq_final_MB})), also weakens the strength of the $N^*(1650)$
because, while $gG$ for this channel is stronger for $N^*(1650)$ than
for $N^*(1535)$ (see Table~\ref{tab_gG}), the present process
cannot be initiated by the $K\Sigma$ channel.}
However, the mass distribution in Fig.~\ref{fig_massdist_1} still has a
larger width
compared to
the mass distribution of the Model~I in
Ref.~\cite{Xie:2017erh}, { where only the $N^*(1535)$ is included following
the work {of} Ref.~\cite{Inoue:2001ip}
using
the chiral unitary approach
without the $\rho N(I=1/2)$ and $\pi \Delta$($d$ wave) channels.}
Meanwhile, the width of the mass distribution of the
$\Lambda_c^+\rightarrow\bar{K}^0\eta p$ is smaller than those of
Models~II, III and IV in Ref.~\cite{Xie:2017erh}.
In these models, {the $N^*(1535)$ is treated as a Breit-Wigner
amplitude and
its}
width is larger than that obtained in
Refs.~\cite{Inoue:2001ip,Garzon:2014ida} or has energy dependence which
makes the width effectively large at higher energy.

For
completeness,
the $K\Sigma$ mass distribution of the
$\Lambda_c^+\rightarrow\bar{K}^0K\Sigma$ decay is shown at the
bottom of Fig.~\ref{fig_massdist_1}.
In Ref.~\cite{Garzon:2014ida}, the value of $g_{N^*,K\Sigma}G_{K\Sigma}$
is larger for the $N^*(1650)$ resonance than the $N^*(1535)$ resonance, and
the $N^*(1535)$ energy is about 200~MeV below the $K\Sigma$ threshold.
Then, we can expect that the $K\Sigma$ production is mainly driven by the
$N^*(1650)$ resonance.
However, as given in Eq.~(\ref{eq_final_MB}) the $K\Sigma$ pair is not
produced directly from the $\Lambda_c$ decay.
Then, the $K\Sigma$ pair is produced only through the coupled channel effect
of the meson-baryon amplitude $t_{MB,M'B'}$ in our approach, and
the magnitude of the mass distribution is much smaller compared with
that of $\pi N$ or $\eta N$.

\begin{figure}[t]
 \centering
 \includegraphics[width=7.5cm]{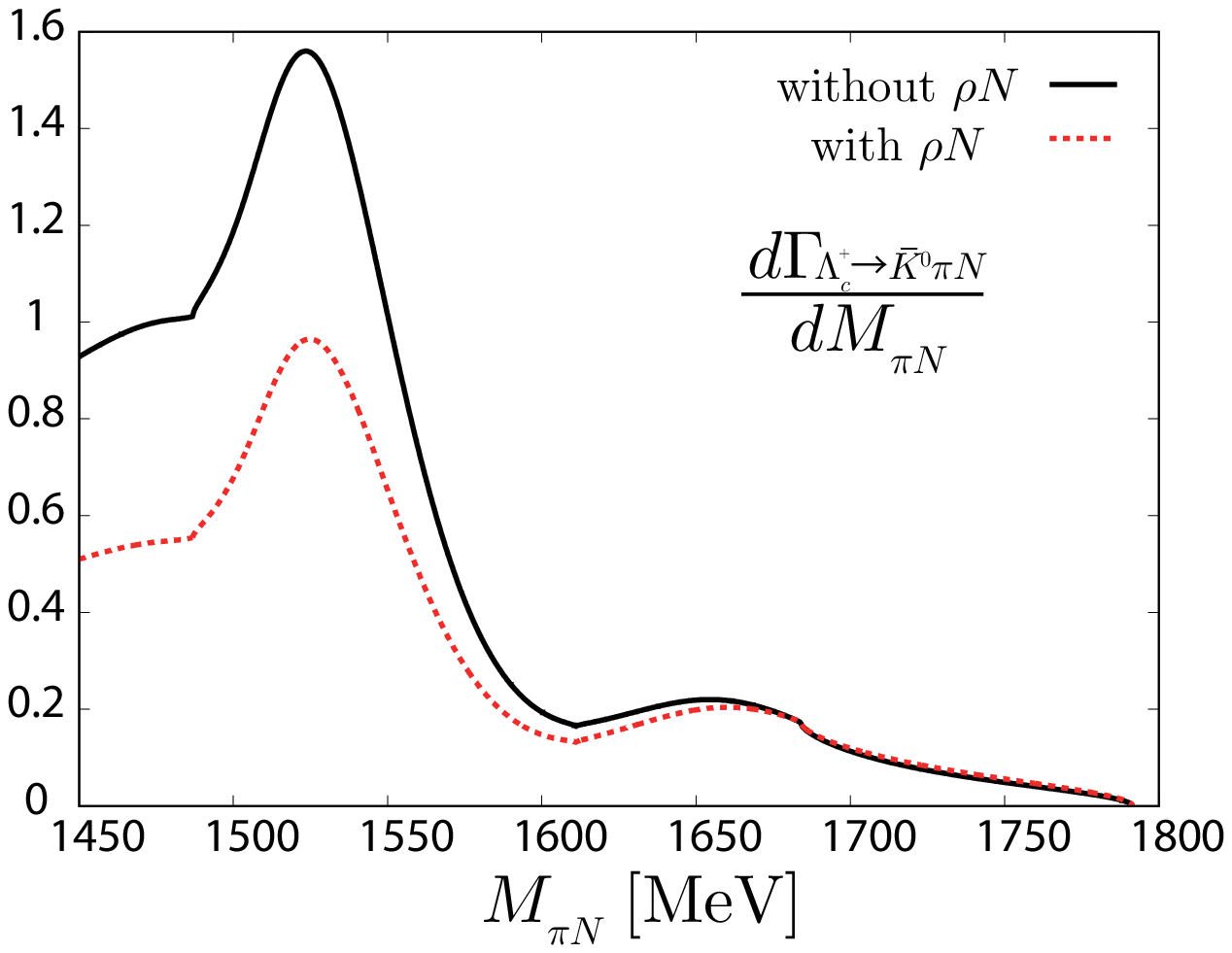}
 \includegraphics[width=7.5cm]{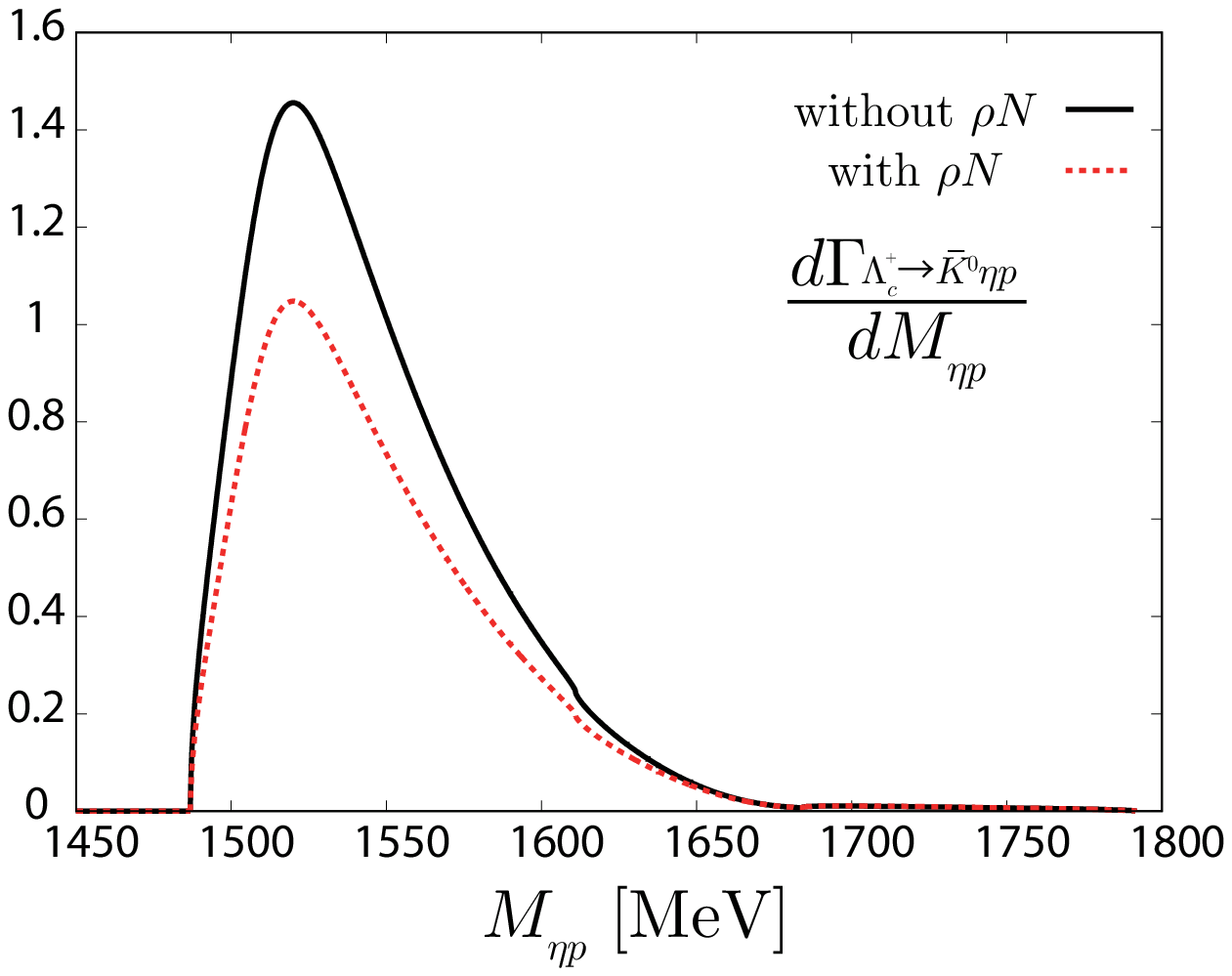}
 \caption{The mass distribution for $\Lambda_c^+\rightarrow\bar{K}^0\pi
 N$ with $I=1/2$ (top) and $\bar{K}^0\eta p$ (bottom) without $\rho N$ channel.
 The mass distributions with the $\rho N$ channel are shown with the
 dotted lines.}
 \label{fig_massdist_no_rhon}
\end{figure}

In Fig.~\ref{fig_massdist_no_rhon}, we show the mass distribution omitting the
$\rho N$ channel in the sum of $M'B'$ in Eq.~(\ref{eq_amp_final}).
The $\rho N$ channel contributes in
a
destructive way to the mass
distribution.
In the $\pi N$ case, the effect of the $\rho N$ channel looks more
significant for the lower peak.
This is because, as shown in Ref.~\cite{Garzon:2014ida}, the $\rho N$
channel has a larger value of $g_{N^*,\rho N}G_{\rho N}$ for the $N^*(1535)$
resonance {than for} the $N^*(1650)$ resonance.

Thus, in the $\Lambda_c^+$ decay into
$\bar{K}^0MB$ $[MB=\pi N(I=1/2),\eta p$ and $K\Sigma(I=1/2)]$ the
resonances $N^*(1535)$ and $N^*(1650)$ appear {in a different} way
than in
the meson-baryon amplitude in Ref.~\cite{Garzon:2014ida}.
In addition, we found a difference from the {models which do not
contain the $N^*(1650)$, or} with respect the
five-quark
models
of the $N^*(1535)$ that {were}
discussed
in Ref.~\cite{Xie:2017erh}.
Then, the production of the $N^*(1535)$ and $N^*(1650)$ from the $\Lambda_c^+$ decay is a good
process to clarify the properties of the $N^*(1535)$ and $N^*(1650)$
resonances.

\section{Summary}
\label{sec_summary}
We have studied the mass distribution of the $\Lambda_c^+\rightarrow
\bar{K}^0MB$ $[MB=\pi N(I=1/2),\eta p$, and $K\Sigma(I=1/2)]$ including the effect of the
$N^*(1535)$ and $N^*(1650)$ resonances which are generated by the
hadron
dynamics with the $\pi N,\eta N,K\Lambda,K\Sigma,\rho N$, and
$\pi\Delta$($d$ wave) channels as investigated in
Ref.~\cite{Garzon:2014ida}.
While
{both}
effects of the $N^*(1535)$ and $N^*(1650)$ are seen in the
mass distributions, we found that their
manifestation is
different from
that
in the meson-baron amplitude given in Ref.~\cite{Garzon:2014ida}, or
experiment.
{In our mass} distribution {for $\Lambda_c^+\rightarrow\bar{K}^0\pi
p(I=1/2)$ and $\bar{K}^0\eta p$}, the peak from $N^*(1535)$ is larger than that
from $N^*(1650)$, while
two peaks with a comparable magnitude are seen
in the
amplitude {of the $\pi N$ to $\pi N$
channel} in Ref.~\cite{Garzon:2014ida}.
This is because the $K\Sigma$ channel which couples more strongly to
$N^*(1650)$ than $N^*(1535)$ is suppressed in the primary production from
$\Lambda_c^+$ in our treatment of the weak and hadronization processes and the $\rho N(I=1/2)$ and $K\Lambda$ channels have larger
couplings to the $N^*(1535)$ resonance than the $N^*(1650)$ resonance.

Furthermore,
we find
differences from
the treatment
of
the $N^*(1535)$ made in
Ref.~\cite{Xie:2017erh}, where
a
{five-quark component of the
$N^*(1535)$ is included using
a
Breit-Wigner amplitude}.

In the case of $MB=K\Sigma(I=1/2)$,
the $N^*(1650)$ resonance is expected to give
a dominant contribution to the production amplitude,
{
but we
found that the magnitude of the mass distribution of the
$\Lambda_c^+\rightarrow\bar{K}^0K\Sigma(I=1/2)$
is much smaller than
for
the other processes, like
$\Lambda_c^+\rightarrow\bar{K}^0\eta p$, because the production of
the $K\Sigma$ is suppressed in the weak and hadronization process.}

The subtleties and results in the different channels in the reactions
studied here are tied to the nature of the $N^*(1535)$ and $N^*(1650)$
resonances as dynamically generated from the hadron interaction in
coupled channels, and the experimental observation of these decay modes
should bring new information concerning the nature of these states.

\section*{Acknowledgements}
 R.P. Pavao wishes to thank the Generalitat Valenciana in the program
 Santiago Grisolia.
 This work is partly supported by the Spanish Ministerio de Economia y
 Competitividad and European FEDER funds under the contract number
 FIS2014-57026-REDT, FIS2014-51948-C2-1-P, and FIS2014-51948-C2-2-P, and
 the Generalitat Valenciana in the program Prometeo II-2014/068.

\bibliographystyle{plain}

\end{document}